\documentclass[manuscript]{aastex}

\usepackage{natbib}
\usepackage{amsmath}

\begin{document}
\def\vector#1{\mbox{\boldmath $#1$}}
\renewcommand{\topfraction}{0.9}
\renewcommand{\bottomfraction}{0.9}
\renewcommand{\dbltopfraction}{0.9}
\renewcommand{\textfraction}{0}
\renewcommand{\floatpagefraction}{0.9}
\title{Nonlinear reflection process of linearly polarized, broadband Alfv\'{e}n waves in the fast solar wind}

\author{M. Shoda and T. Yokoyama}
\affil{Department of Earth and Planetary Science, The University of Tokyo, Bunkyo-ku, Tokyo 113-0033, Japan}
\email{shoda@eps.s.u-tokyo.ac.jp}

\begin{abstract}
Using one-dimensional numerical simulations, we study the elementary process of Alfv\'{e}n wave reflection in a uniform medium, including nonlinear effects.
In the linear regime, Alfv\'{e}n wave reflection is triggered only by the inhomogeneity of the medium, whereas in the nonlinear regime, it can occur via nonlinear wave-wave interactions.
Such nonlinear reflection (backscattering) is typified by decay instability.
In most studies of decay instabilities, the initial condition has been a circularly polarized Alfv\'{e}n wave. 
In this study we consider a linearly polarized Alfv\'en wave, which drives density fluctuations by its magnetic pressure force.
For generality, we also assume a broadband wave with a red-noise spectrum.
In the data analysis, we decompose the fluctuations into characteristic variables using local eigenvectors, thus revealing the behaviors of the individual modes.
Different from circular-polarization case, we find that the wave steepening produces a new energy channel from the parent Alfv\'en wave to the backscattered one.
Such nonlinear reflection explains the observed increasing energy ratio of the sunward to the anti-sunward Alfv\'{e}nic fluctuations in the solar wind with distance against the dynamical alignment effect.
\end{abstract}
\keywords{magnetohydrodynamics(MHD) --- methods: numerical --- Sun: corona --- Sun: solar wind--- Sun: Alfv\'{e}n wave}

\section{Introduction}
Alfv\'{e}n waves are frequently observed in the solar atmosphere \citep{DeP07,Oka07}.
As Alfv\'en waves upwardly transport large amounts of photospheric kinetic energy \citep{Fuj09}, they are considered promising sources of coronal heating and acceleration of fast solar wind in coronal holes.
In this framework, rapid dissipation of Alfv\'{e}n waves should inject thermal energy into the corona and accelerate the solar wind via wave pressure \citep{Dew70,Bel71b,Hol73,Jac77,Hei80}.
As the magnetic flux tubes expand in the coronal hole region, the nonlinearity of the Alfv\'{e}n waves become significantly large so that nonlinear interactions can initiate an energy cascade, playing a significant role in the dissipation process. 

Nonlinear dissipation of Alfv\'en waves can occur by two main candidate mechanisms: compressible and incompressible processes.
In the first mechanism, nonlinear coupling between Alfv\'{e}n and compressible waves initiate an energy cascade by steepening.
Unless they are monochromatic and circularly polarized, Alfv\'{e}n waves can generate acoustic waves via their magnetic pressure forces \citep{Hol71}, which then steepen into shock waves.
Alfv\'{e}n waves can also directly steepen to form fast (switch-on) shocks or rotational discontinuities \citep{Mon59,Coh74,Ken90}.
This steepening and mode conversion process are suggested to explain many solar phenomena such as spicule formation \citep{Hol82,Kud99,Mat10}, coronal heating \citep{Mor04,Suz05,Suz06,Ant08,Mat14}, and solar wind acceleration \citep{Suz05,Suz06,Mat14}.

The second mechanism is Alfv\'{e}nic turbulence triggered by two bidirectional Alfv\'{e}n waves \citep{Iro64,Kra65,Goldr95}.
Due to nonuniform Alfv\'en speed in coronal holes, Alfv\'{e}n waves generated by photospheric motion are linearly reflected \citep{Fer58,An90,Vel93,Cra05,Hol07}.
The resulting interaction between the original anti-sunward and reflected sunward waves triggers Alfv\'enic turbulence.
This turbulence model is also suggested as a candidate for coronal heating and solar wind acceleration \citep{Cra07,Ver10} using a phenomenological turbulent dissipation mode \citep{Hos95,Dmi02,Cha09}.
As shown in recent three-dimensional direct numerical simulations using a reduced magnetohydrodynamics (MHD) approximation, Alfv\'{e}nic turbulence can heat both closed coronal loops \citep{Bal11} and open coronal holes \citep{Woo15}.

The amount of reflected wave energy flux is of critical importance in the Alfv\'enic turbulence models.
Most studies on turbulence-driven coronal heating and solar wind acceleration have assumed locally incompressible \citep{Zho90} or nearly incompressible \citep{Zan92} plasma.
However, Alfv\'{e}n waves can be reflected by nonlinear interactions with compressible waves.
The best-known example of such nonlinear reflection is the decay instability \citep{Sag69,Gol78,Hos89,Zan01}.
In addition, \citet{Suz05} investigated a compressible heating/acceleration model and reported significant reflection of Alfv\'en waves.
Therefore the role of compressibility in reflection triggering should not be discounted.
To more realistically model the corona and solar wind, it is necessary to elucidate the elementary processes of Alfv\'{e}n wave reflection caused by nonlinear interactions with compressible modes.
This problem has motivated the study in this paper.

\citet{Bel71a} reported that solar wind fluctuations near Earth's orbit are mainly composed of anti-sunward Alfv\'{e}n waves.
Later studies revealed that the ratio of sunward to anti-sunward Alfv\'{e}n wave energies increases with distance \citep{Bav00}.
This fact is against the fundamental process of Alfv\'enic turbulence called dynamical alignment \citep{Dob80}.
To explain this trend, researchers have proposed theoretical models based on the compressible turbulence \citep{Gra93}, linear reflection \citep{Ver07} and decay instability \citep{Zan01}.

Herein, we propose a new reflection process that explains this observational fact.
This process is a combination of the wave steepening and the decay instability, that is, linear polarization and resultant steepening forms a new energy channel which never appears in the circular-polarization case. 
We show that our result with a proper scaling well agree with the observation by \citet{Bav00}.

The remainder of this paper is organized as follows.
In $\S$ 2 we describe the numerical setting and data analysis method.
The results of the numerical calculation and data analysis are presented in $\S$ 3.
In the last section, $\S$ 4, we propose an elementary reflection process that can explain our results and discuss on the comparison with the observation.

\section{Method}

\subsection{Numerical setting}

For simplicity, we assume a one-dimensional Cartesian coordinate system.
Periodic boundary conditions are used.
We denote by $x$ the spatial coordinate, background magnetic field $\vector{B}_0$ parallel to the $x$ axis.
Because the sole transverse component is the $y$ component, the Alfv\'{e}n waves are linearly polarized.
We also assume a static background with a uniform density $\rho_0$ and uniform magnetic field $B_0$.
The initial state can then expressed as 
\begin{equation}
\rho=\rho_0, \ \ {v_x}=0, \ \ {v_y}=0, \ \ {B_x}=B_0, \ \ {B_y}=0.
\end{equation}
The dissipation mechanisms, such as viscosity and resistivity, are not explicitly stated.
For brevity, we further assume an isothermal system with speed of sound $C_s$.
The basic equations of the system are thus given by
\begin{equation}
\frac{\partial}{\partial t} \rho + \frac{\partial }{\partial x} \left( \rho v_x \right) =0,
\end{equation}
\begin{equation}
\frac{\partial}{\partial t} \left( \rho v_x \right) + \frac{\partial }{\partial x} \left( \rho {v_x}^2 + \rho {C_s}^2 + \frac{{B_y}^2}{8\pi} \right)=0,
\end{equation}
\begin{equation}
\frac{\partial}{\partial t} \left( \rho v_y \right) + \frac{\partial }{\partial x} \left( \rho v_x v_y - \frac{B_0 B_y}{4 \pi} \right)=0,
\end{equation}
\begin{equation}
\frac{\partial}{\partial t} B_y + \frac{\partial }{\partial x} \left( B_y v_x - B_0 v_y \right)=0.
\end{equation}

The equations are numerically solved by using an upwind scheme with the linearized Riemann solver (Roe's solver) developed for isothermal MHD systems by \citet{Nak96} and \citet{Fuk99}.
The spatial and temporal accuracies in this scheme are set to be second-order, and unphysical numerical oscillations near the discontinuities are avoided by a minmod flux limiter.

\subsection{Initial condition as the wave input}

We input the waves by the initial condition.
The fluctuations are denoted by $\Delta$ and initially exist as a purely rightward Els\"{a}sser state without fluctuations in $\rho$ or $v_x$:
\begin{equation}
\Delta \rho=0, \ \ \Delta {v_x}=0, \ \ \Delta {v_y}=C_A f(x), \ \ \Delta {B_y}=-B_0 f(x),
\end{equation}
where $C_A = B_0 / \sqrt{4 \pi \rho_0}$ is the background Alfv\'{e}n speed, and $f(x)$ represents the initial wave profile.
Notice that if $f(x)$ is sufficiently small, the leftward Els\"{a}sser variable vanishes, and the initial condition becomes a purely rightward linear Alfv\'{e}n wave.

The initial fluctuation is assumed to have a red-noise energy spectrum with random phase.
Previous observations \citep{Mat10} revealed that within a certain band (typically with period between 1 min and 10 min), the energy spectrum of photospheric transverse motion approximates a red-noise spectrum.
By setting the highest wavenumber of the initial fluctuation spectrum equal to one tenth of the Nyquist wavenumber, itself given as one-half of the spatial grid points $N_x$, $f(x)$ can be explicitly written as
\begin{equation}
f(x)= \ \sum^N_{n=1}  A_n \sin \left[ 2 \pi \left( \frac{n x}{L} \right) + \phi_n \right]
\end{equation}
where $N=N_x/20$, $A_n=A_0 n^{-1}$ and $L$ is the size of the simulation box.
$A_0$ is the amplitude parameter, and $\phi_n$ is a random value ranging between $0$ and $2 \pi$.
We perform two kinds of simulations: short-term simulations up to $t=20 \tau_A$ with $N_x=5000$ and long-term evolutions up to $t=1000 \tau_A$ with $N_x=1000$ where $\tau_A = L/C_A$ is the Alfv\'{e}n time of the simulation box.
Our simulation is characterized by two free physical parameters: the plasma beta $\beta=(C_s/C_A)^2$ and the initial fluctuation energy $E_{wave}$, which is defined as follows:
\begin{eqnarray}
E_{wave}
&=& \int^L_0 dx \left[ \frac{1}{2} \rho_0 {\Delta {v_y}}^2 + \frac{1}{8 \pi} {\Delta {B_y}}^2 \right] \nonumber \\
&=& \frac{1}{2} \rho_0 (A_0 C_A)^2 L  \sum^N_{n=1} n^{-2}.
\end{eqnarray}

\subsection{Decomposition of fluctuations into characteristic variables
\label{sec:met:dec}}

To understand the nonlinear evolution in terms of normal modes of MHD, we adopt the decomposition of the fluctuations into characteristic variables.
In an isothermal system, wave dissipation does not increase the temperature (and the speed of sound); consequently the background (mean) field is always steady and uniform, and the mean and fluctuation of each variable are easily decoupled. 

In terms of primitive variables $\vector{W}^{\mathrm T}=(\rho, v_x, v_y, B_y)$, the governing equations, Eqs. $(2)-(5)$, can be rewritten as
\begin{equation}
\frac{\partial }{\partial t} \vector{W} + \vector{A} \left( \vector{W} \right) \cdot \frac{\partial}{\partial x}\vector{W} = 0,
\end{equation}
where $\vector{A} \left( \vector{W} \right)$ is the characteristic matrix of the primitive variables, explicitly written as
\begin{equation}
  \vector{A} \left( \vector{W} \right) = \left(
    \begin{array}{cccc}
      v_x & \rho & 0 & 0 \\
      {C_s}^2/\rho & v_x & 0 & B_y/\rho \\
      0 & 0 & v_x & -B_0/\rho \\
	0 & B_y & -B_0 & v_x
    \end{array}
  \right).
\end{equation}
Following \citet{Sto08}, the right and left eigenmatrices of $\vector{A} \left( \vector{W} \right)$ are explicitly given as
\begin{equation}
  \vector{R} \left( \vector{W} \right) = \left(
    \begin{array}{cccc}
      \rho \alpha_f & \rho \alpha_s & \rho \alpha_s & \rho \alpha_f \\
      -C_{ff} & -C_{ss} & C_{ss} & C_{ff} \\
      C_{ss} & -C_{ff} & C_{ff} & -C_{ss} \\		
      A_s & -A_f & -A_f & A_s
    \end{array}
  \right),
\end{equation}
and
\begin{equation}
  \vector{L} \left( \vector{W} \right) = \left(
    \begin{array}{cccc}
      \alpha_f /2 \rho & - C_{ff}/2 {C_s}^2 &  C_{ss}/2 {C_s}^2 & A_s /2 \rho {C_s}^2 \\
      \alpha_s /2 \rho & - C_{ss}/2 {C_s}^2 & - C_{ff}/2 {C_s}^2 & - A_f /2 \rho {C_s}^2 \\
      \alpha_s /2 \rho & C_{ss}/2 {C_s}^2 &  C_{ff}/2 {C_s}^2& - A_f /2 \rho {C_s}^2 \\
	\alpha_f /2 \rho & C_{ff}/2 {C_s}^2 & - C_{ss}/2 {C_s}^2 & A_s /2 \rho {C_s}^2
    \end{array}
  \right),
\end{equation}
where each variable is a function of the local fast and slow mode velocities $C_{fast}$ and $C_{slow}$, respectively:
\begin{equation*}
{C_{fast}}^2 = \frac{1}{2} \left[ {C_s}^2 + \frac{{B_0}^2+{B_y}^2}{4 \pi \rho} + \sqrt{\left({C_s}^2 + \frac{{B_0}^2+{B_y}^2}{4 \pi \rho} \right)^2 - 4 {C_s}^2 \frac{{B_0}^2}{4 \pi \rho}} \hspace{0.5em} \right],
\end{equation*}
\begin{equation*}
{C_{slow}}^2 = \frac{1}{2} \left[ {C_s}^2 + \frac{{B_0}^2+{B_y}^2}{4 \pi \rho} - \sqrt{\left({C_s}^2 + \frac{{B_0}^2+{B_y}^2}{4 \pi \rho} \right)^2 - 4 {C_s}^2 \frac{{B_0}^2}{4 \pi \rho}} \hspace{0.5em} \right],
\end{equation*}
\begin{equation*}
{\alpha}_f=\frac{{C_s}^2-{C_{slow}}^2}{{C_{fast}}^2-{C_{slow}}^2}, \ \ {\alpha}_s=\frac{{C_{fast}}^2-{C_s}^2}{{C_{fast}}^2-{C_{slow}}^2},
\end{equation*}
\begin{equation*}
C_{ff}=C_{fast} {\alpha}_f, \ \ C_{ss}=C_{slow} {\alpha}_s,
\end{equation*}
\begin{equation*}
A_f=C_s {\alpha}_f\sqrt{\rho} \ \ \mbox{and} \ \ A_s=C_s {\alpha}_s \sqrt{\rho}.
\end{equation*}
The mean field $\vector{W}_0$ is trivial:
\begin{equation}
	\vector{W}_0 =
\left(
    \begin{array}{c}
      \rho_0 \\
      0 \\
      0 \\
      0
    \end{array}
  \right),
\end{equation}
and the fluctuation field can easily be obtained by subtracting $\vector{W}_0$ from $\vector{W}$ as $\Delta \vector{W} = \vector{W}-\vector{W}_0$.
The fluctuation can then be decomposed via ``the local'' right eigenmatrix  $\vector{R} \left( \vector{W} \right) \left( \neq \vector{R}(\vector{W}_0 ) \right)$ as 
\begin{equation}
\Delta \vector{W} =\vector{R} \left( \vector{W} \right) \cdot \vector{\alpha} \left( \vector{W} \right)
\label{decompose-a}
\end{equation}
or more explicitly,
\begin{equation}
\left(
    \begin{array}{c}
      \Delta \rho \\
      \Delta v_x \\
      \Delta v_y \\
      \Delta B_y
    \end{array}
  \right)	
= \left(
    \begin{array}{cccc}
      \rho \alpha_f & \rho \alpha_s & \rho \alpha_s & \rho \alpha_f \\
      -C_{ff} & -C_{ss} & C_{ss} & C_{ff} \\
      C_{ss} & -C_{ff} & C_{ff} & -C_{ss} \\		
      A_s & -A_f & -A_f & A_s
    \end{array}
  \right)
\left(
    \begin{array}{c}
      \alpha_{fl} \\
      \alpha_{sl} \\
      \alpha_{sr} \\
      \alpha_{fr} 
    \end{array}
  \right),
\label{eq:decompose-a}
\end{equation}
where $\vector{\alpha}$ is a vector of amplitudes.
In terms of $\vector{\alpha}$, this equation can be solved as
\begin{equation}
\vector{\alpha} \left( \vector{W} \right) =\vector{L} \left( \vector{W} \right) \cdot \Delta \vector{W}.
\end{equation}
Each column of matrix $\vector{R}$ corresponds to a right eigenvector of $\vector{A}$: ${\vector{\psi}}_{i}$.
Therefore, we can rewrite $\vector{R}$ as $\vector{R}=\left( \vector{\psi}_{fl}, \vector{\psi}_{sl}, \vector{\psi}_{sr}, \vector{\psi}_{fr} \right)$.
The fluctuation of mode $i$, $\Delta \vector{W}_{i}$ is the following product of $\alpha_i$ and ${\vector{\psi}}_{i}$
\begin{equation}
\Delta \vector{W}_{i} = \alpha_i \vector{\psi}_{i} \ \ \left( i = fl, sl, sr, fr \right) 
\end{equation}
where $fl$ and $sl$ denote leftward Alfv\'{e}n and acoustic waves, respectively, and $sr$ and $fr$ denote the corresponding rightward waves.

\section{Results}

\subsection{Time evolutions of density, velocity and magnetic field}

\subsubsection{Evolutions of raw data}

\begin{figure}[htbp]
\begin{flushright}
\includegraphics[width=150mm]{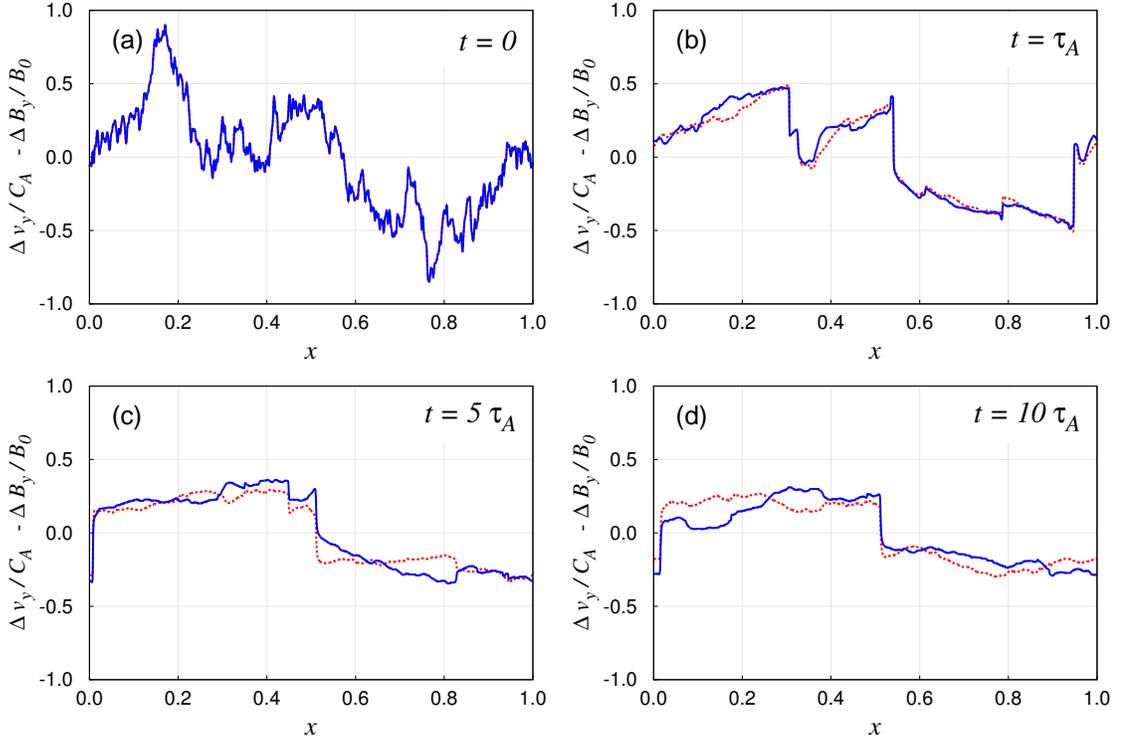} \ \ 
\end{flushright}
\caption{
Snapshots of $\Delta v_y / C_A$ (blue, solid line) and $- \Delta B_y / B_0$ (red, dotted line) in typical case ($E_{wave} / E_{gas} =0.5 \ \ C_s / C_A =0.5$).
Snapshots are captured at (a) $t=0$, (b) $t=\tau_A$, (c) $t=5 \tau_A$ and (d) $t=10 \tau_A$.
}
\vspace{1em}
\label{vely}
\end{figure}

\begin{figure}[htbp]
\begin{flushright}
\includegraphics[width=150mm]{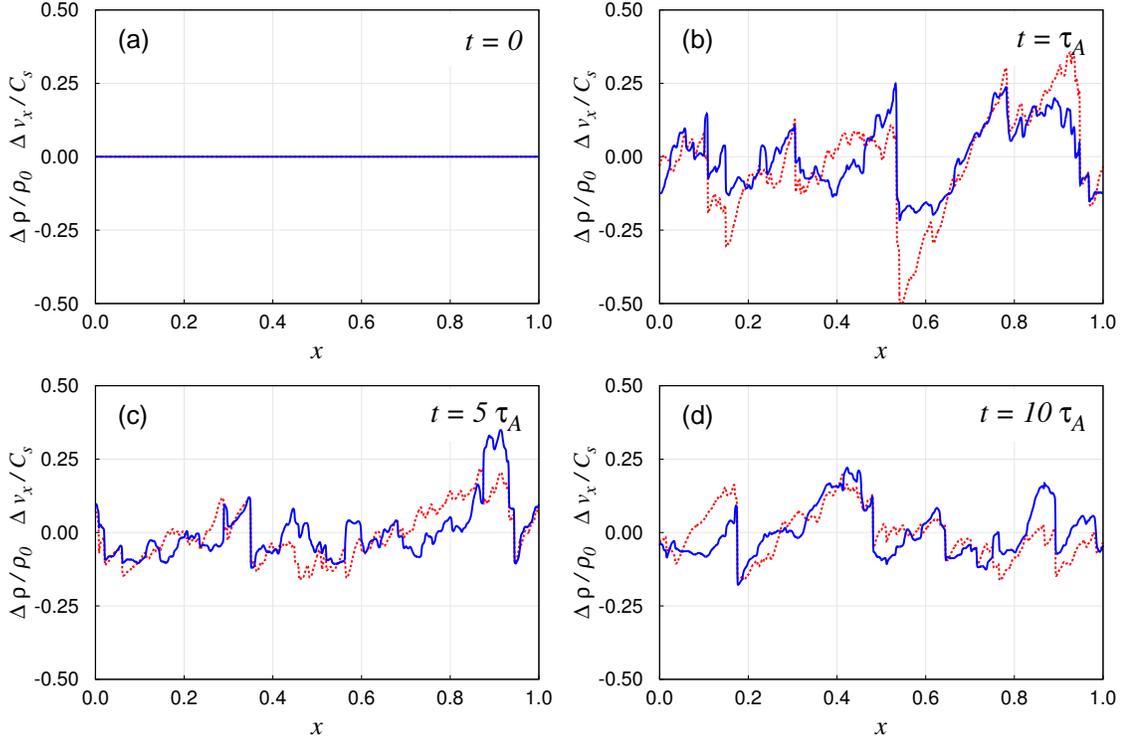} \ \ 
\end{flushright}
\caption{
Snapshots of $\Delta \rho / \rho_0$ (blue, solid line) and $\Delta v_x / C_s$ (red, dotted line).
Parameters and evolution times are those of Fig. \ref{vely}.
}
\vspace{1em}
\label{dens}
\end{figure}

Figures \ref{vely} and \ref{dens} show the time evolutions of normalized physical quantities in a typical case in which the initial wave energy $E_{wave}$ equals one-half of the background thermal energy $E_{gas}$:
\begin{equation}
E_{gas}= \rho_0 {C_s}^2 L,
\end{equation}
The background plasma beta is set to $\beta=0.25 \ (C_s/C_A=0.5)$.
Fig. \ref{vely} presents the evolutions of transverse fluctuations $\Delta v_y, \ \Delta B_y$.
Steepening of the magnetic field is seen, which has been analytically explained by \citet{Mon59}, \citet{Coh74} and \citet{Ken90}.
The decreased number of shock fronts results not only from dissipation but also from the merging of shocks.
In other words, when two fast shock waves collide, they merge into a stronger fast shock.
In this case, a few strong shocks successively overtake many weak shocks and merge with them, reducing their number.
This apparent inverse cascade should not be interpreted as the energy transport toward large scales.
The deviation of $\Delta v_y / C_A$ from $-\Delta B_y / B_0$ in Fig. \ref{vely} confirms that reflection occurs in our simulation, because it shows that the nonzero component of the normalized leftward Els\"{a}sser variable $z^{-}=\Delta v_y / C_A + \Delta B_y / B_0$ exists.

Fig. \ref{dens} reveals many slow (acoustic) shock waves generated by the magnetic pressure of wave.
This effect is called nonlinear mode conversion from Alfv\'{e}n waves to acoustic waves \citep{Hol82,Kud99,Mor04,Suz05}.
Unlike in the transverse field, the number of shock fronts does not evidently decrease even at $t=10\tau_A$.
There are several reasons for this.
First is the long overtaking time of the slow shock waves caused by the small speed of sound (in the present low-beta case).
Second, the shock formations occur at different time scales.
The formation times of fast and slow shock waves are inversely proportional to the square of the nonlinearity and the  nonlinearity itself, respectively.
That is, denoting the shock formation times of fast and slow waves by $\tau_f$ and $\tau_s$, respectively, we have
\begin{equation}
\tau_f \propto \left( \frac{\Delta v_y}{C_A} \right)^{-2} \ \ \ \ \tau_s \propto \left( \frac{\Delta v_x}{C_s} \right)^{-1}.
\end{equation}
Since the relation $\Delta v_x / C_s \sim \Delta v_y / C_A < 1$ is generally satisfied in our calculation, slow shock waves are more easily generated than fast shock waves.
Third, new acoustic waves are continuously generated by the magnetic pressure of the Alfv\'{e}n waves.
Despite the shock dissipation, the amplitude decreases less markedly than in transverse fields, which supports the third reason

We also perform the simulations with initially monochromatic wave with a wavelength of the box size $L$.
Although there are several differences, the evolutions were similar with the red-noise case, especially for the transverse fields.
This is because the red-noise wave has the largest energy in the longest-wavelength mode.

\subsubsection{Evolutions of decomposed data}

\begin{figure}[htbp]
\begin{flushright}
\includegraphics[width=150mm]{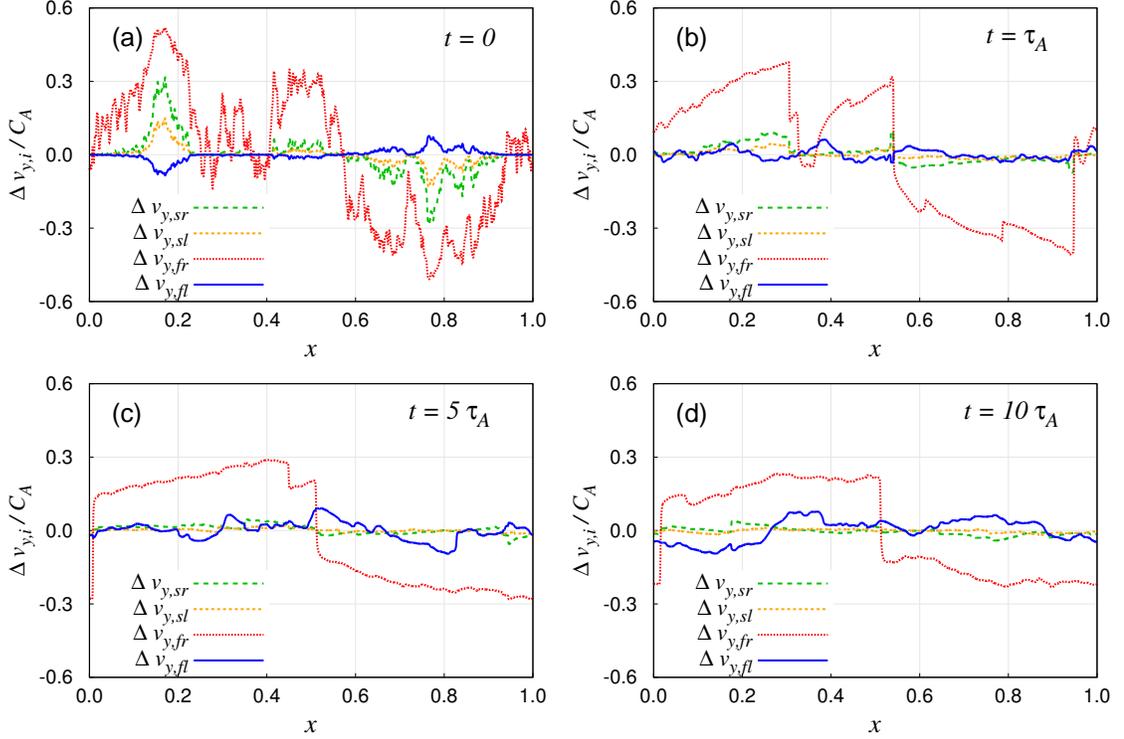} \ \ 
\end{flushright}
\caption{
Results of decomposed fluctuation $\Delta v_{y,i}$.
Four lines correspond to modes $\Delta v_{y,fr}$ (red, dotted line), $\Delta v_{y,fl}$ (blue, solid line), $\Delta v_{y,sr}$ (green, long-dashed line) and $\Delta v_{y,sl}$ (orange, short-dashed line).
}
\vspace{1em}
\label{vely-decomp}
\end{figure}
\begin{figure}[htbp]
\begin{flushright}
\includegraphics[width=150mm]{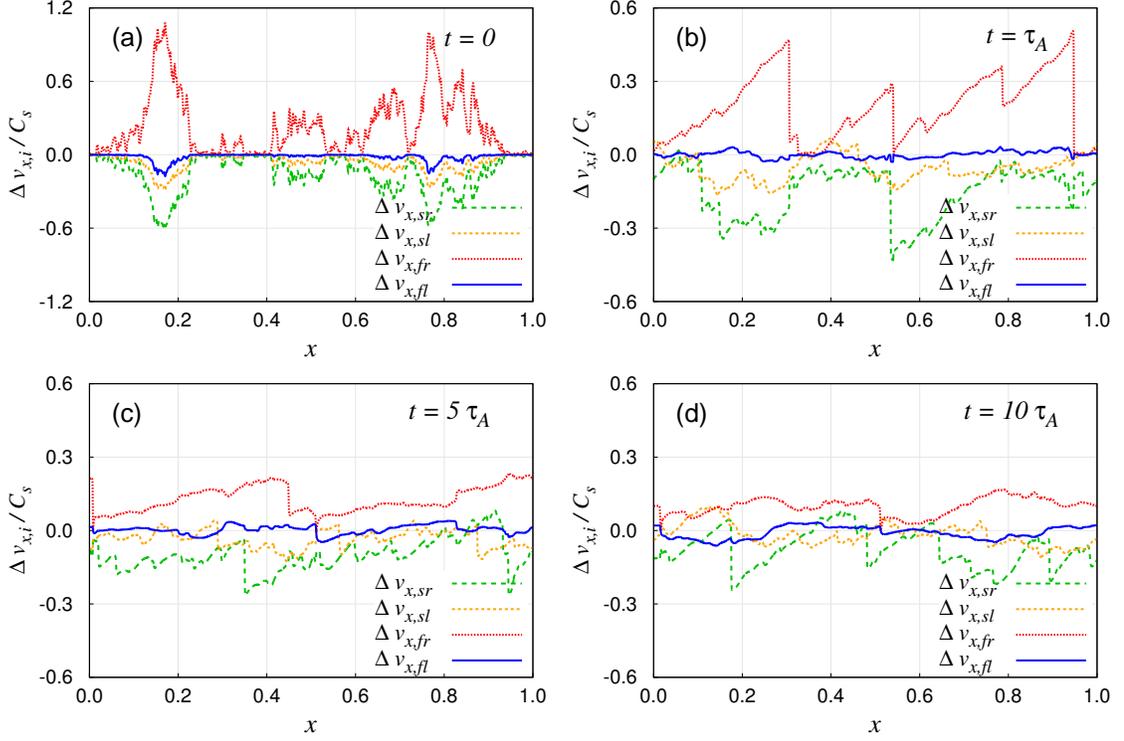} \ \ 
\end{flushright}
\caption{
Results of decomposed fluctuation $\Delta v_{x,i}$.
Four lines correspond to modes $\Delta v_{x,fr}$ (red, dotted line), $\Delta v_{x,fl}$ (blue, solid line), $\Delta v_{x,sr}$ (green, long-dashed line) and $\Delta v_{x,sl}$ (orange, short-dashed line).
Vertical axis scale differs between (a) and (b).
}
\vspace{1em}
\label{velx-decomp}
\end{figure}

Figures \ref{vely-decomp} and \ref{velx-decomp} show the evolutions in the typical case of the decomposed fluctuations $\Delta v_{y,i}$ and $\Delta v_{x,i}$ (where $i$ represents a mode) defined in Section \ref{sec:met:dec}.
Although we initially impose a purely rightward Els\"{a}sser state, not only a rightward Alfv\'{e}nic fluctuation $\Delta v_{y,fr}$, but also those of other modes, $\Delta v_{y,fl}$, $\Delta v_{y,sr}$, $\Delta v_{y,sl}$ appear (Fig. \ref{vely-decomp}).
This contamination originates from the finite-amplitude effect, which deviates the Els\"{a}sser variables from the fast-mode characteristics.
The most important feature in this figure is the increase of the amplitude of $\Delta v_{y,fl}$ from $t=0$ to $t=10 \tau_A$, which provides direct evidence of reflection.

Fig. \ref{velx-decomp} shows the behavior of the longitudinal fluctuations.
It is initially transported by the rightward Alfv\'{e}n wave.
However, the amplitude of $\Delta v_x$ gradually shifts from the rightward Alfv\'{e}n wave to the rightward acoustic (slow) wave.
This shift occurs probably because nonlinear interactions of Alfv\'{e}n waves decrease the nonlinearity of the Alfv\'en waves and generate acoustic waves.

\subsection{Self-energy evolution}

\subsubsection{Short-term behavior}

In this subsection, we discuss the physical mechanism of wave reflection.
For this purpose, we examine the evolutions of the individual modes.
In terms of the decomposed density $\Delta \rho_i$, velocities $\Delta v_{x,i}$ and $\Delta v_{y,i}$, and the magnetic field $\Delta B_{y,i}$, the normalized ``self-energy'' of each wave mode $i$ can be defined as
\begin{equation}
E_i = \int_0^L dx \left[ \frac{1}{2} \left( \rho_0 + \Delta \rho_i \right) \left( {\Delta {v_{x,i}}}^2 + {\Delta {v_{y,i}}}^2 \right) + \frac{{\Delta {B_{y,i}}}^2}{8\pi} \right] / E_{wave} \ \ \left( i = fl, sl, sr, fr \right).
\end{equation}
Note that the summation of $E_i$ over modes $i$ does not agree with the total energy of the fluctuations because it excludes the interaction terms between two modes.

\begin{figure}[htbp]
\begin{flushright}
\includegraphics[width=120mm]{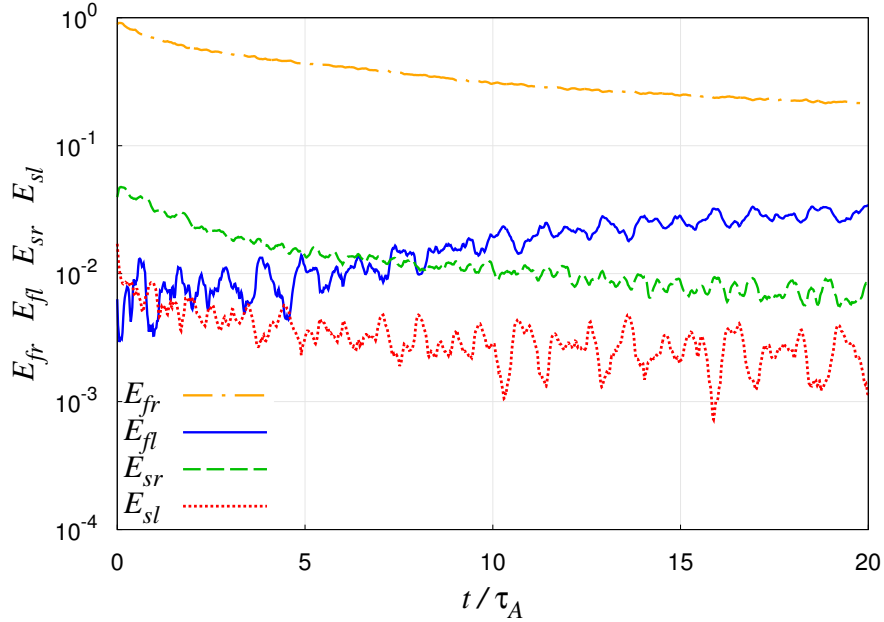} \ \ \ \ \ \ \ \ \ \ \ \ \ \ 
\end{flushright}
\caption{
Time evolutions of self-energies in the typical case ($E_{wave} / E_{gas}=0.5$ and $C_s / C_A=0.5$).
Four lines represent rightward Alfv\'{e}n wave energy $E_{fr}$ (orange, dash-dotted line), leftward Alfv\'{e}n wave energy $E_{fl}$ (blue, solid line), rightward acoustic wave energy $E_{sr}$ (green, dashed line), and leftward acoustic wave energy $E_{sl}$ (red, dotted line).
}
\vspace{1em}
\label{0.5-1}
\end{figure}

Figure \ref{0.5-1} shows the evolving self-energies in the typical case $(E_{wave} / E_{gas} =0.5, \ C_s / C_A = 0.5)$.
Quasi-periodic energy oscillations are evident in each mode. 
In particular, the phases of the leftward Alfv\'{e}n wave energy $E_{fl}$ and leftward acoustic wave energy $E_{sl}$ are negatively correlated.
This phase-anticorrelated oscillation suggests the exchange of energy via resonance between these two modes.
Other simulation runs, in which we changed $E_{wave} / E_{gas}$ and $C_s / C_A$ (results not shown), confirm that phase-anticorrelated oscillations always appear except such cases with high wave energy ($E_{wave}>E_{gas}$) and extremely low plasma beta.

\begin{figure}[htbp]
\begin{flushright}
\includegraphics[width=120mm]{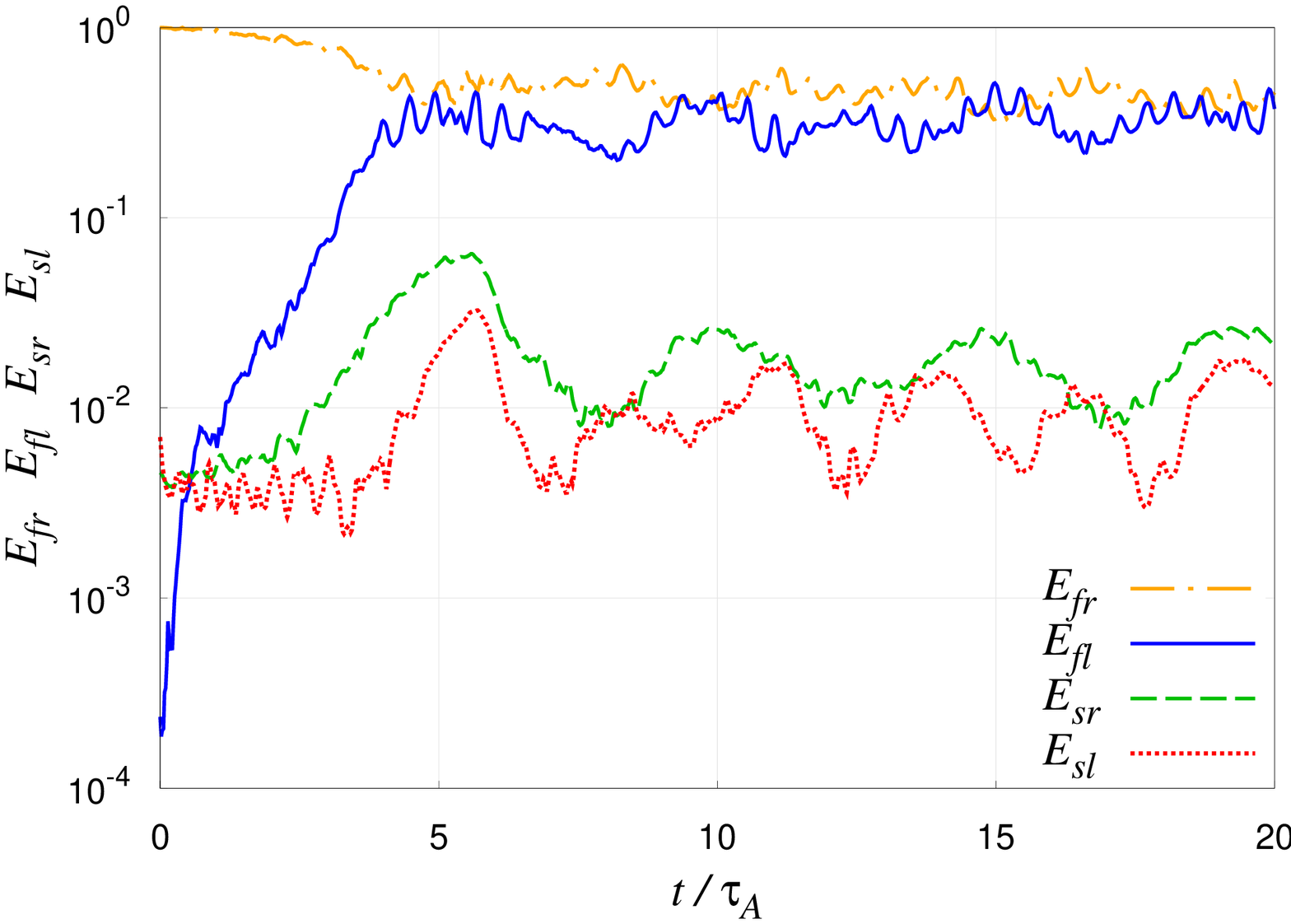} \ \ \ \ \ \ \ \ \ \ \ \ \ \ 
\end{flushright}
\caption{
Decay instability case.
Parameters are set to $C_s/C_A=0.1$ and $E_{wave}/E_{gas}=2$.
}
\vspace{1em}
\label{decay}
\end{figure}

Figure \ref{decay} shows the results of a case with the low plasma beta and the large amplitude ($C_s / C_A=0.1$ and $E_{wave} / E_{gas}=2$).
We have found that the decay instability appears.
This finding is important, as it confirms the possibility of decay instability in linearly polarized waves.
The extremely-low beta condition is essential, because decay instability never occurs when $C_s/C_A=0.5$.
Note that the evolutions of self-energies (especially of the leftward Alfv\'en wave) in Fig. \ref{decay} differs from those in Fig. \ref{0.5-1}, indicating that the physical process commonly observed in our simulations differs from the usual decay instability.
$E_{fl}$ increases almost exponentially between $t=\tau_A$ and $t=4\tau_A$, providing direct evidence of some instability.
Since this is an instability of Alfv\'en waves in low-beta plasmas, it is definitely the decay instability.
$E_{fl}$ saturates at values comparable to $E_{fr}$, when the feedback process is no longer negligible.
This feedback appears in Fig. \ref{decay} after $t=5\tau_A$, when $E_{fr}$ increases by the feedback from the leftward Alfv\'en waves.

To examine the wave energy oscillation and correlation between two modes in detail, we average each $E_i$ over time.
The averaged energy here is defined as 
\begin{equation}
\overline{E_i}(t)=\sqrt[2N_{A}+1]{\prod^{N_{A}}_{j=-N_{A}} E_i(t+j\Delta t)},
\end{equation}
where $\Delta t$ is the cadence of the data and $N_{A} \Delta t$ is the Afv\'{e}n time ($N_A = \tau_A/\Delta t$).
The fluctuation part $\Delta E_i$ is defined in terms of $\overline{E_i}$ as
\begin{equation}
\Delta E_i = E_i - \overline{E_i}.
\end{equation}
To illustrate the dominant correlation between the leftward Alfv\'en and acoustic modes, we generate scatter plots between the energy fluctuations of different modes  $\Delta E_i$ and $\Delta E_j$ and computed the correlation factors $C$.
Both of them reveal the strongest correlation between $\Delta E_{fl}$ and $\Delta E_{sl}$ among combinations of the modes.
From these facts we inferred the successive exchange of energy, mainly between the leftward Alfv\'{e}n and acoustic waves.

\begin{figure}[htbp]
\vspace{-10truemm} 
\begin{flushright}
\includegraphics[width=150mm]{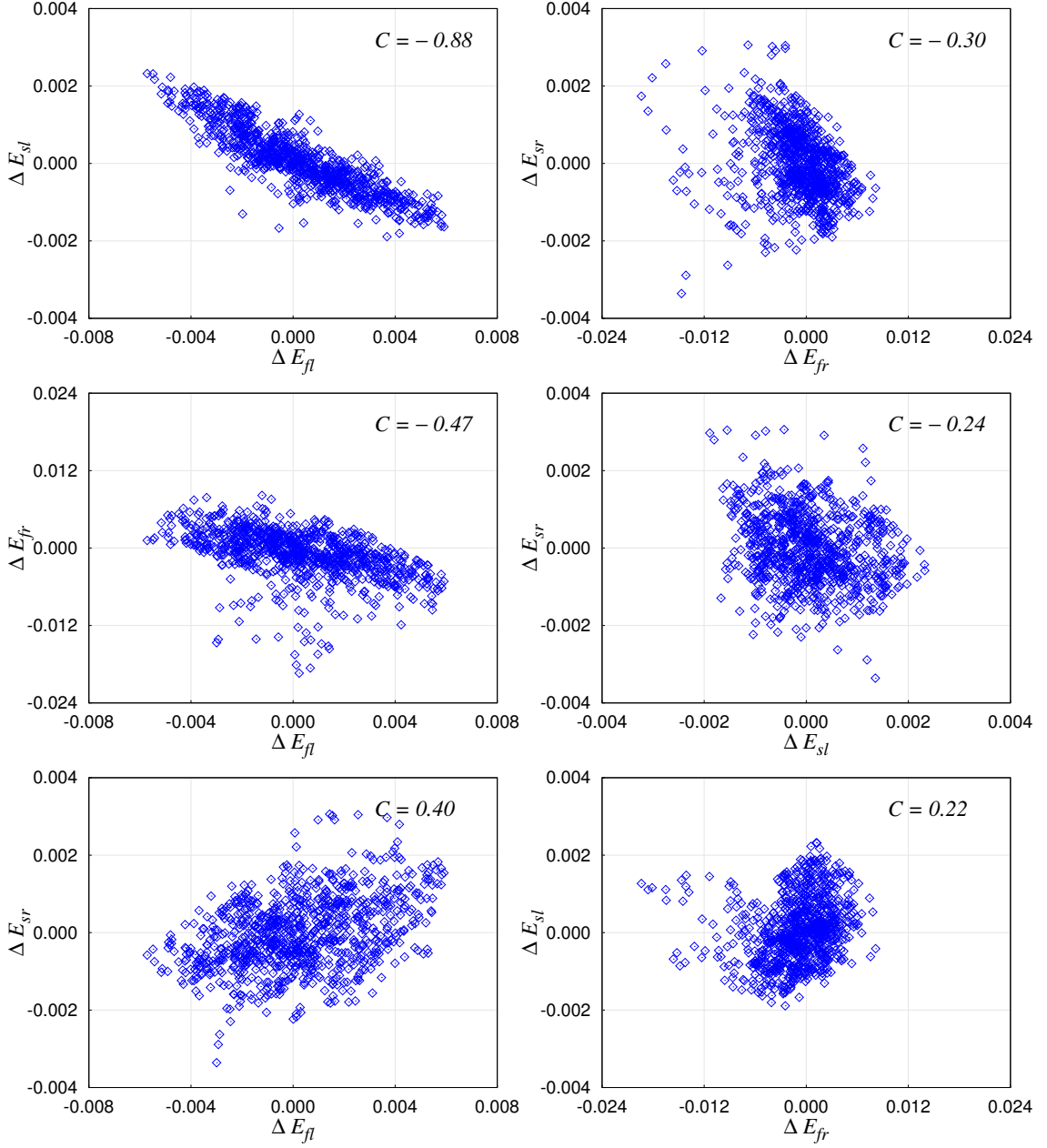} \ \ 
\end{flushright}
\caption{
Scatter plots for all $\Delta E_i$ and $\Delta E_j$.
Correlation factor $C$ of each set is displayed inside corresponding panel.
As demonstrated by plot profiles and $C$ values, strongest correlation occurs between $\Delta E_{fl}$ and $\Delta E_{sl}$.
We have confirmed this result in other parameter sets.
}
\vspace{1em}
\label{scatter}
\end{figure}

\subsubsection{Long-term behavior}

\begin{figure}[htbp]
\begin{flushright}
\includegraphics[width=120mm]{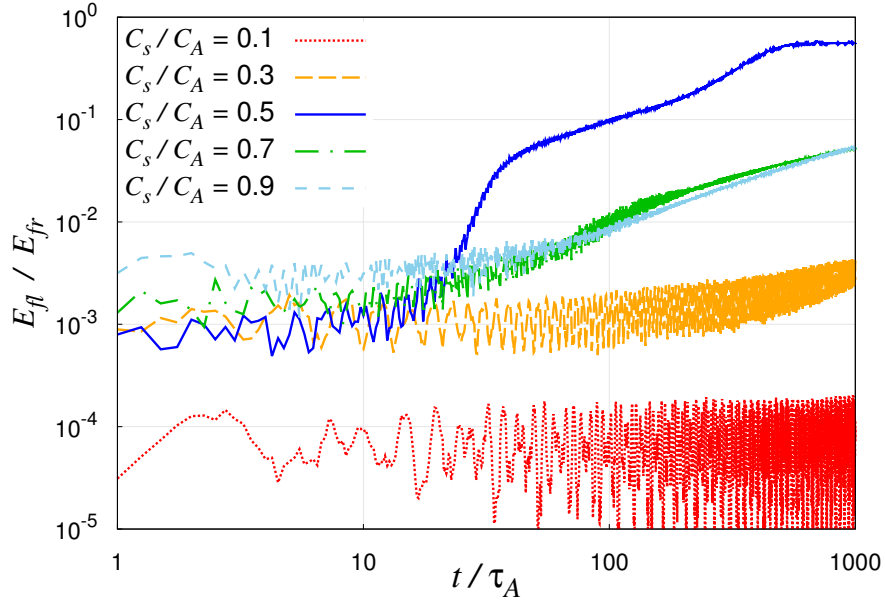} \ \ \ \ \ \ \ \ \ \ \ \ \ \ 
\end{flushright}
\caption{
Evolutions of population ratio with same energy ratio $(E_{wave}/E_{gas}=0.25)$ and different plasma betas $C_s/C_A=0.1$ (red, dotted line), $0.3$ (orange, long-dashed line), $0.5$ (blue, solid line), $0.7$ (green, dash-dotted line), $0.9$ (light-blue, short-dashed line).
}
\vspace{1em}
\label{log2^-2}
\end{figure}

We now present the long-term evolution of the system, up to $t=1000 \tau_A$.
In particular, we focus on the population ratio (or reflection ratio) defined as $E_{fl}/E_{fr}$. 
In Fig. \ref{log2^-2}, it is shown for cases with $C_s/C_A = 0.1, 0.3, 0.5, 0.7,$ or $0.9$, fixing $E_{wave} / E_{gas} = 0.25$.
When $C_s / C_A =0.5$, the population ratio is anomalously rapidly enhanced around $t=20\tau_A$.
This efficient reflection is an essentially nonlinear phenomenon that amplifies the growth rate over time.
When the initial energy of the fluctuations equals the thermal energy of the background, that is,  when $E_{wave}=E_{gas}$, fast reflections occur not only when $C_s/C_A=0.5$ but also when $C_s/C_A=0.7$ and $C_s/C_A=0.9$.
The increased nonlinearity permits rapid enhancement of the population ratio over wider parameter space, because sufficient energy is available for the reflection.
Such rapid enhancement also supports that the wave resonance is critical, because the timescale of the resonance decreases with increasing amplitude of the coupled waves.

\section{Discussion}

\subsection{Physical process of reflection}

In the previous section, we demonstrate two enhancement features of $E_{fl}$ and $E_{fl}/E_{fr}$.

1. Nonmonotonic increase of the leftward Alfv\'en wave energy $E_{fl}$, which involves short-period oscillations anti-correlated with the leftward acoustic wave energy $E_{sl}$.

2. Nonconstant (initially increasing) growth of the population ratio $E_{fl}/E_{fr}$.
The temporal evolution of $E_{fl}/E_{fr}$ is nonmonotonically sensitive to the $\beta$ value. 
In particular, when $C_s/C_A=0.5$, the population ratio is enhanced at an anomalous rate, as shown in Fig. \ref{log2^-2}.

In this section, we discuss theoretically an elementary reflection process that explains above features.
We focus on the typical case ($E_{wave} / E_{gas} =0.5, C_s / C_A =0.5$) here.
To clarify the physical process of the reflection, we perform Fourier transformation in space to the normalized velocity fluctuation $\Delta v_{y,fr}$, $\Delta v_{y,fl}$, $\Delta v_{x,sr}$ and $\Delta v_{x,sl}$.
Notice that normalization factors are different between fast (Alfv\'en) and slow (acoustic) modes.
We refer to each Fourier mode by the wavenumber normalized by $2 \pi / L$, that is, mode $p$ represents a mode whose wavenumber is $2 \pi p / L$ (wave length is $L /p$).
Due to the periodic boundary condition, $p$ is limited to integers.

First, we focus on feature 1, i.e., phase-anticorrelated energy oscillations between two leftward waves.
Resonance between these two waves can naturally explain this behavior, that is, interaction between some parent and leftward Alfv\'{e}n waves generates a leftward acoustic mode via three-wave resonance and similarly, interaction between some parent and a leftward acoustic waves generates a leftward Alfv\'{e}n wave.
This is a stable process because, since the wave momenta should be conserved during wave-wave interactions, leftward Alfv\'en and acoustic waves cannot amplify their energies at the same time but just oscillate their energies.
Rightward Alfv\'en waves can be a parent wave (i.e., energy mediator) in this process, because, different from the decay instability, three-wave resonance with a forward Alfv\'en wave as a parent wave and backward Alfv\'en and acoustic waves as daughter waves is a stable process.
This resonant energy exchange may play an important role in the saturation of $E_{fl}$, because $E_{fl}$ is likely to be transported to $E_{sl}$, which is much easier to dissipate via the shock wave dissipation than $E_{fl}$.

\begin{figure}[htbp]
\begin{flushright}
\includegraphics[width=177mm]{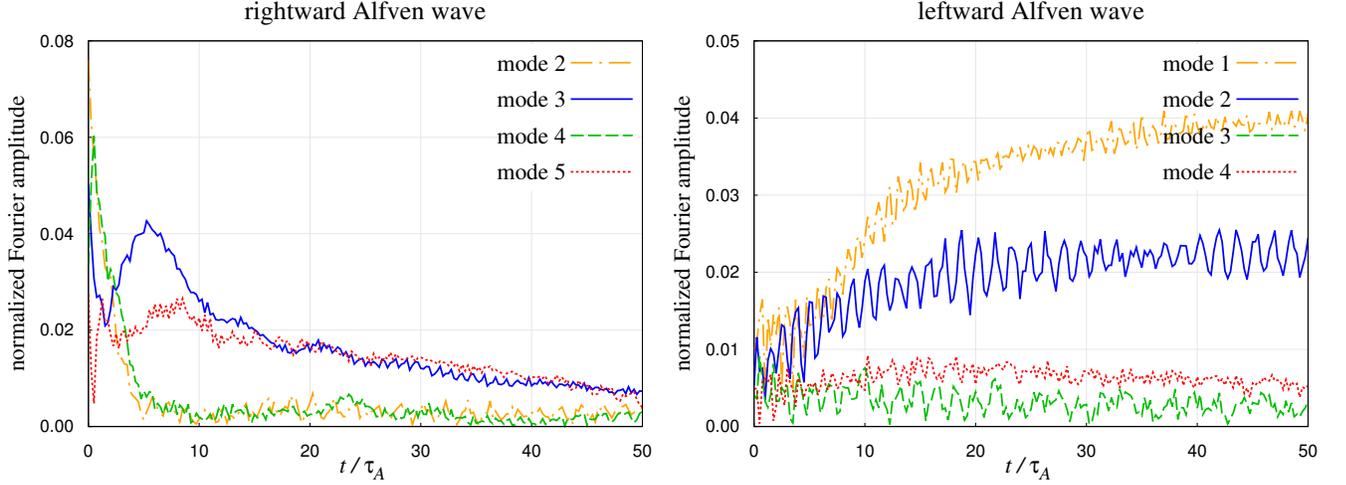}
\end{flushright}
\caption{Time evolutions of the Fourier amplitudes of $\Delta v_{y,fr} / C_A$ and $\Delta v_{y,fl} / C_A$ the for typical case $(E_{wave}/E_{gas}=0.5, C_s/C_A=0.5)$.
Left panel shows the evolutions of mode 2 (orange, dash-dotted line), mode 3 (blue, solid line), mode 4 (green, dashed line) and mode 5 (red, dotted line) of rightward Alfv\'en waves, while right panel shows those of mode 1 (orange, dash-dotted line), mode 2 (blue, solid line), mode 3 (green, dashed line) and mode 4 (red, dotted line) of leftward Alfv\'en waves, respectively.
}
\label{fast}
\vspace{1em}
\end{figure}

Next, we discuss on the amplification process of $E_{fl}$.
In Figure \ref{fast}, we show the Fourier-transformed amplitudes of $\Delta v_{y,fr}$ and $\Delta v_{y,fl}$.
In the left panel of Fig. \ref{fast}, where rightward Alfv\'en wave is focused, we show the temporal evolutions of modes 2 to 5.
Although all modes decrease their amplitudes initially, odd-number modes recover their amplitude.
This is caused by the steepening of mode 1:
It generates density fluctuations (not acoustic waves) of mode 2 via the magnetic pressure and cascades (steepens) by interacting with the density fluctuations.
Interaction between mode 1 Alfv\'en wave and mode 2 density fluctuation generates mode 3 Alfv\'en wave, which as well generates mode 5 Alfv\'en wave by the interaction with the density fluctuation.
In this way, only odd-number modes survive.

\begin{figure}[htbp]
\begin{center}
\includegraphics[width=130mm]{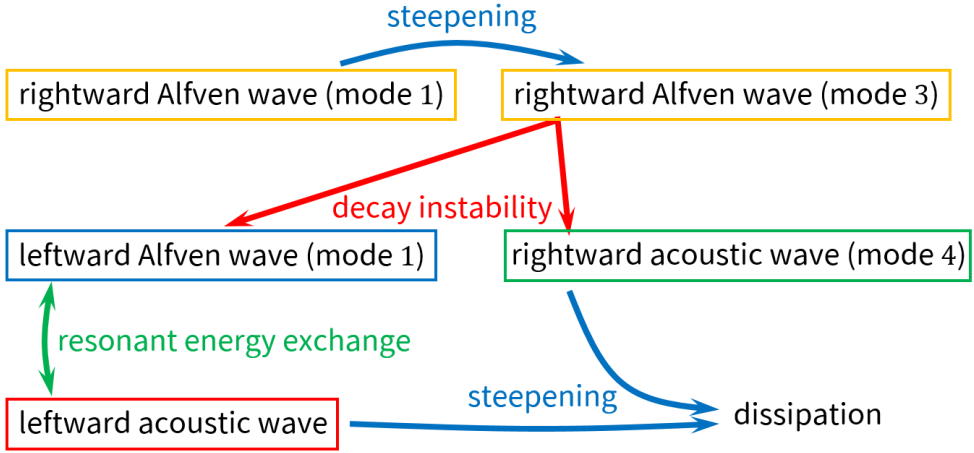} \ \ \ \ \ \ \ \ 
\end{center}
\caption{
Schematic diagram of the energy channels found in the typical case.
}
\label{ponti}
\vspace{1em}
\end{figure}

After the steepening, lower-number modes tend to have higher amplitudes, which is not the case in Fig. \ref{fast} after $t=17\tau_A$.
This behavior indicates some energy absorption from mode 3 rightward Alfv\'en wave.
In the right panel of Fig. \ref{fast} we show the leftward Alfv\'en waves of modes 1 to 4.
The dominant mode of the reflected Alfv\'en wave is mode 1.
Considering the possibility of some energy absorption from mode 3 rightward Alfv\'en wave, the decay instability is the promising process of the amplification, because mode 3 rightward Alfv\'en and mode 1 leftward Alfv\'en waves can satisfy the three-wave resonance condition of the decay instability in this plasma beta ($C_s / C_A =0.5$).
\begin{figure}[htbp]
\begin{flushright}
\includegraphics[width=120mm]{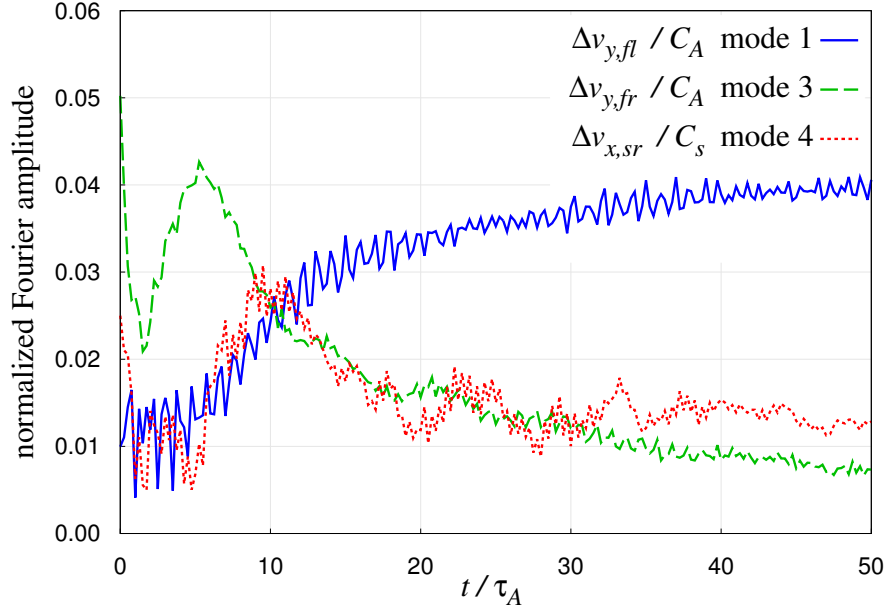} \ \ \ \ \ \ \ \ \ \ \ \ \ \ 
\end{flushright}
\caption{
Evidence of the decay instability.
Each line shows the velocity amplitude of mode 1 leftward Alfv\'en wave (blue, solid line), mode 3 rightward Alfv\'en wave (green, dashed line) and mode 4 rightward acoustic wave (red, dotted line).
}
\label{decfou}
\vspace{1em}
\end{figure}
This is explained as follows.
The three-wave resonance condition of the decay instability is given as
\begin{equation}
\left(
    \begin{array}{c}
      k_0 \\
      \omega_0
    \end{array}
  \right)
= 
\left(
    \begin{array}{c}
      - k_1 \\
      \omega_1
    \end{array}
  \right) 
+
\left(
    \begin{array}{c}
      k_2 \\ 
      \omega_2
    \end{array}
  \right)
\label{eq:3wave}
\end{equation}
\begin{equation}
\omega_0 = k_0 C_A \ \ \ \ \omega_1 = -k_1 C_A \ \ \ \ \omega_2 = k_2 C_s, \nonumber
\end{equation}
where $k_0$, $k_1$ and $k_2$ are wavenumbers of the rightward Alfv\'en (parent), leftward Alfv\'en and rightward acoustic waves, respectively.
After solving this, we obtain following relation.
\begin{equation}
\left( k_0 : k_1 : k_2 \right) = \left( 1+\frac{C_s}{C_A} : 1-\frac{C_s}{C_A} : 2 \right).
\label{eq:decaypair}
\end{equation}
By substituting $C_s / C_A =0.5$, it is shown that mode 1 leftward Alfv\'en wave can absorb the energy of mode 3 rightward Alfv\'en wave via the decay instability with mode 4 rightward acoustic wave another daughter wave.
To support this, we show in Figure \ref{decfou} Fourier-transformed velocity amplitudes of these three modes, that is, mode 3 rightward Alfv\'en, mode 1 leftward Alfv\'en and mode 4 rightward acoustic waves.
It is clear that leftward Alfv\'en and rightward acoustic waves increase their energy at the same time as the rightward Alf\'en wave decays.
We have confirmed that the energy increase of mode 4 rightward acoustic wave is not by the steepening, because it exceeds energies of the other modes when it reaches its maximum.
If circularly polarized, mode 1 Alfv\'en waves never steepen but decay directly to mode 1 backscattered Alfv\'en waves, which is the usual process of the decay instability.
This difference shows that linear polarization and resultant steepening generate a new energy channels, which is effective at least in the typical case.
In Figure \ref{ponti}, we summarize these new-found elementary processes and energy channels by a schematic diagram.

\subsection{Application to open system}

\begin{figure}[htbp]
\begin{flushright}
\includegraphics[width=150mm]{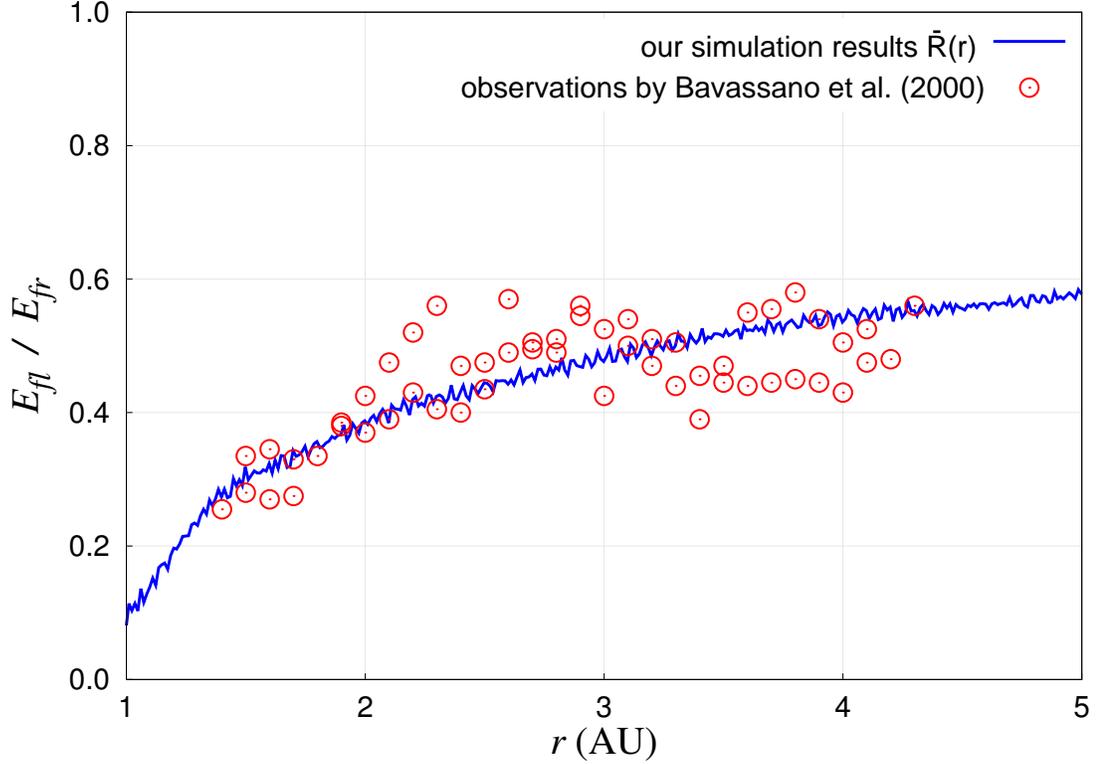}
\end{flushright}
\caption{
Energy ratios of leftward and rightward propagating Alfv\'en waves: Continuous curve and open circles denote our simulation results ($\overline{R}$, see text for details), and observations by \citet{Bav00}, respectively.
}
\label{compare}
\vspace{1em}
\end{figure}

In this subsection, we present the comparison of our results on the population ratio $E_{fl}/E_{fr}$ with solar wind observations.
Because the plasma beta in the solar wind changes with distance, some technical interpretation is necessary.
Our results and the observational data are not directly comparable.
Instead, we average our results for various plasma betas.
Specifically, we calculate five cases ($C_s/C_A = 0.5,0.6,0.7,0.8,0.9$) with fixed the initial energy ratio $E_{wave} / E_{gas}=1$, which is observationally appropriate according to \citet{Golds95}.
We then average the results by using Eq. (\ref{eq:averageR}).
Here $R(t;\alpha)$ and $\overline{R}(t)$ denote the time evolution of $E_{fl}/E_{fr}$ with $C_s/C_A=\alpha$ and the averaged result, respectively.
\begin{equation}
\overline{R}(t) = \frac{1}{5} \left[ R(t;0.5) + R(t;0.6) + R(t;0.7) + R(t;0.8) + R(t;0.9) \right].
\label{eq:averageR}
\end{equation}
Next we convert time in our results to the distance in real solar wind. 
We use
\begin{equation}
r-r_{0.1}=V_{SW} (t-t_{0.1})
\label{eq:scaling}
\end{equation}
in this conversion, where $V_{SW}$ is the solar wind speed assumed as $750$ km/sec, $r$ is the radius from the sun center, $t_{0.1}$ is the time at which $E_{fl}/E_{fr}$=0.1 in our results, and $r_{0.1}$ is the location at which $\overline{R}(t)=0.1$ in the fast solar wind observations.
According to \citet{Golds95}, $r_{0.1}=1$ AU.
The time on the right hand side of Eq. (\ref{eq:scaling}) is normalized by the Alfv\'{e}n time $\tau_A$, assigned as the dominant wave period in the solar wind fluctuations, i.e., $10^4$ sec.
This value is determined from the dominant frequency ($10^{-4}$ Hz) in the measured fast solar wind fluctuations \citep{Tu95}.
The obtained $\overline{R}(r)$ are compared with the observational values of \citet{Bav00} in Figure \ref{compare}.
Our results favorably agree with the observations.
To our knowledge, these trends have been best explained by linear reflection \citep{Ver07} and decay instability \citep{Zan01}.
However, the linear reflection process requires the most dominant frequency of Alfv\'{e}n waves to be  $10^{-6}$ Hz.
At Alfv\'en wave frequencies around $10^{-4}$ Hz, linear reflection cannot generate sufficient reflected waves enough \citep{Cra05}.
On the other hand, decay instability cannot explain the saturation of $E_{fl}/E_{fr}$ around $0.5$ beyond 3 AU.
Our results (Fig. \ref{compare}) explain the observations while alleviating both difficulties.

There are several points to improve which are simplified in this study.
First, since our system is periodic in space, we need to confirm that our proposed process operates in open systems.
Second, we should consider kinetic effects in the solar wind condition; most critically, Landau damping of (ion-)acoustic waves.
When the Landau damping is too strong, the acoustic waves dissipate before the backscattering via the decay instability.
Third, one-dimensional uniform-background simulations usually overestimate the shock effects and neglect the expansion effect of the solar wind \citep{Gra93,Nar15,Zan15}.
For an investigation of this influence, it requires multidimensional simulations.

\acknowledgments
The authors acknowledge Takeru K. Suzuki and Takuma Matsumoto for their fruitful discussions.
One of the authors (M. Shoda) also thanks Masahiro Hoshino for his discussions and critical advice.
This research is supported by Leading Graduate Course for Frontiers of Mathematical Sciences and Physics (FMSP) and JSPS KAKENHI Grant Number 15H03640.


\begin{thebibliography}{53}
\expandafter\ifx\csname natexlab\endcsname\relax\def\natexlab#1{#1}\fi

\bibitem[{{An} {et~al.}(1990){An}, {Suess}, {Moore}, \& {Musielak}}]{An90}
{An}, C.-H., {Suess}, S.~T., {Moore}, R.~L., \& {Musielak}, Z.~E. 1990, \apj,
  350, 309

\bibitem[{{Antolin} {et~al.}(2008){Antolin}, {Shibata}, {Kudoh}, {Shiota}, \&
  {Brooks}}]{Ant08}
{Antolin}, P., {Shibata}, K., {Kudoh}, T., {Shiota}, D., \& {Brooks}, D. 2008,
  \apj, 688, 669

\bibitem[{{Bavassano} {et~al.}(2000){Bavassano}, {Pietropaolo}, \&
  {Bruno}}]{Bav00}
{Bavassano}, B., {Pietropaolo}, E., \& {Bruno}, R. 2000, \jgr, 105, 15959

\bibitem[{{Belcher}(1971)}]{Bel71b}
{Belcher}, J.~W. 1971, \apj, 168, 509

\bibitem[{{Belcher} \& {Davis}(1971)}]{Bel71a}
{Belcher}, J.~W., \& {Davis}, Jr., L. 1971, \jgr, 76, 3534

\bibitem[{{Chandran} {et~al.}(2009){Chandran}, {Quataert}, {Howes}, {Hollweg},
  \& {Dorland}}]{Cha09}
{Chandran}, B.~D.~G., {Quataert}, E., {Howes}, G.~G., {Hollweg}, J.~V., \&
  {Dorland}, W. 2009, \apj, 701, 652

\bibitem[{{Cohen} \& {Kulsrud}(1974)}]{Coh74}
{Cohen}, R.~H., \& {Kulsrud}, R.~M. 1974, Physics of Fluids, 17, 2215

\bibitem[{{Cranmer} \& {van Ballegooijen}(2005)}]{Cra05}
{Cranmer}, S.~R., \& {van Ballegooijen}, A.~A. 2005, \apjs, 156, 265

\bibitem[{{Cranmer} {et~al.}(2007){Cranmer}, {van Ballegooijen}, \&
  {Edgar}}]{Cra07}
{Cranmer}, S.~R., {van Ballegooijen}, A.~A., \& {Edgar}, R.~J. 2007, \apjs,
  171, 520

\bibitem[{{De Pontieu} {et~al.}(2007){De Pontieu}, {McIntosh}, {Carlsson},
  {Hansteen}, {Tarbell}, {Schrijver}, {Title}, {Shine}, {Tsuneta}, {Katsukawa},
  {Ichimoto}, {Suematsu}, {Shimizu}, \& {Nagata}}]{DeP07}
{De Pontieu}, B., {et~al.} 2007, Science, 318, 1574

\bibitem[{{Del Zanna} {et~al.}(2015){Del Zanna}, {Matteini}, {Landi},
  {Verdini}, \& {Velli}}]{Zan15}
{Del Zanna}, L., {Matteini}, L., {Landi}, S., {Verdini}, A., \& {Velli}, M.
  2015, Journal of Plasma Physics, 81, 013202

\bibitem[{{Del Zanna} {et~al.}(2001){Del Zanna}, {Velli}, \&
  {Londrillo}}]{Zan01}
{Del Zanna}, L., {Velli}, M., \& {Londrillo}, P. 2001, \aap, 367, 705

\bibitem[{{Dewar}(1970)}]{Dew70}
{Dewar}, R.~L. 1970, Physics of Fluids, 13, 2710

\bibitem[{{Dmitruk} {et~al.}(2002){Dmitruk}, {Matthaeus}, {Milano}, {Oughton},
  {Zank}, \& {Mullan}}]{Dmi02}
{Dmitruk}, P., {Matthaeus}, W.~H., {Milano}, L.~J., {Oughton}, S., {Zank},
  G.~P., \& {Mullan}, D.~J. 2002, \apj, 575, 571

\bibitem[{{Dobrowolny} {et~al.}(1980){Dobrowolny}, {Mangeney}, \&
  {Veltri}}]{Dob80}
{Dobrowolny}, M., {Mangeney}, A., \& {Veltri}, P. 1980, Physical Review
  Letters, 45, 144

\bibitem[{{Ferraro} \& {Plumpton}(1958)}]{Fer58}
{Ferraro}, C.~A., \& {Plumpton}, C. 1958, \apj, 127, 459

\bibitem[{{Fujimura} \& {Tsuneta}(2009)}]{Fuj09}
{Fujimura}, D., \& {Tsuneta}, S. 2009, \apj, 702, 1443

\bibitem[{{Fukuda} \& {Hanawa}(1999)}]{Fuk99}
{Fukuda}, N., \& {Hanawa}, T. 1999, \apj, 517, 226

\bibitem[{{Goldreich} \& {Sridhar}(1995)}]{Goldr95}
{Goldreich}, P., \& {Sridhar}, S. 1995, \apj, 438, 763

\bibitem[{{Goldstein}(1978)}]{Gol78}
{Goldstein}, M.~L. 1978, \apj, 219, 700

\bibitem[{{Goldstein} {et~al.}(1995){Goldstein}, {Roberts}, \&
  {Matthaeus}}]{Golds95}
{Goldstein}, M.~L., {Roberts}, D.~A., \& {Matthaeus}, W.~H. 1995, \araa, 33,
  283

\bibitem[{{Grappin} {et~al.}(1993){Grappin}, {Velli}, \& {Mangeney}}]{Gra93}
{Grappin}, R., {Velli}, M., \& {Mangeney}, A. 1993, Physical Review Letters,
  70, 2190

\bibitem[{{Heinemann} \& {Olbert}(1980)}]{Hei80}
{Heinemann}, M., \& {Olbert}, S. 1980, \jgr, 85, 1311

\bibitem[{{Hollweg}(1971)}]{Hol71}
{Hollweg}, J.~V. 1971, \jgr, 76, 5155

\bibitem[{{Hollweg}(1973)}]{Hol73}
---. 1973, \jgr, 78, 3643

\bibitem[{{Hollweg} \& {Isenberg}(2007)}]{Hol07}
{Hollweg}, J.~V., \& {Isenberg}, P.~A. 2007, Journal of Geophysical Research
  (Space Physics), 112, 8102

\bibitem[{{Hollweg} {et~al.}(1982){Hollweg}, {Jackson}, \& {Galloway}}]{Hol82}
{Hollweg}, J.~V., {Jackson}, S., \& {Galloway}, D. 1982, \solphys, 75, 35

\bibitem[{{Hoshino} \& {Goldstein}(1989)}]{Hos89}
{Hoshino}, M., \& {Goldstein}, M.~L. 1989, Physics of Fluids B, 1, 1405

\bibitem[{{Hossain} {et~al.}(1995){Hossain}, {Gray}, {Pontius}, {Matthaeus}, \&
  {Oughton}}]{Hos95}
{Hossain}, M., {Gray}, P.~C., {Pontius}, Jr., D.~H., {Matthaeus}, W.~H., \&
  {Oughton}, S. 1995, Physics of Fluids, 7, 2886

\bibitem[{{Iroshnikov}(1964)}]{Iro64}
{Iroshnikov}, P.~S. 1964, \sovast, 7, 566

\bibitem[{{Jacques}(1977)}]{Jac77}
{Jacques}, S.~A. 1977, \apj, 215, 942

\bibitem[{{Kennel} {et~al.}(1990){Kennel}, {Blandford}, \& {Wu}}]{Ken90}
{Kennel}, C.~F., {Blandford}, R.~D., \& {Wu}, C.~C. 1990, Physics of Fluids B,
  2, 253

\bibitem[{{Kraichnan}(1965)}]{Kra65}
{Kraichnan}, R.~H. 1965, Physics of Fluids, 8, 1385

\bibitem[{{Kudoh} \& {Shibata}(1999)}]{Kud99}
{Kudoh}, T., \& {Shibata}, K. 1999, \apj, 514, 493

\bibitem[{{Matsumoto} \& {Shibata}(2010)}]{Mat10}
{Matsumoto}, T., \& {Shibata}, K. 2010, \apj, 710, 1857

\bibitem[{{Matsumoto} \& {Suzuki}(2014)}]{Mat14}
{Matsumoto}, T., \& {Suzuki}, T.~K. 2014, \mnras, 440, 971

\bibitem[{{Montgomery}(1959)}]{Mon59}
{Montgomery}, D. 1959, Physical Review Letters, 2, 36

\bibitem[{{Moriyasu} {et~al.}(2004){Moriyasu}, {Kudoh}, {Yokoyama}, \&
  {Shibata}}]{Mor04}
{Moriyasu}, S., {Kudoh}, T., {Yokoyama}, T., \& {Shibata}, K. 2004, \apjl, 601,
  L107

\bibitem[{{Nakajima} \& {Hanawa}(1996)}]{Nak96}
{Nakajima}, Y., \& {Hanawa}, T. 1996, \apj, 467, 321

\bibitem[{{Nariyuki}(2015)}]{Nar15}
{Nariyuki}, Y. 2015, Physics of Plasmas, 22, 022309

\bibitem[{{Okamoto} {et~al.}(2007){Okamoto}, {Tsuneta}, {Berger}, {Ichimoto},
  {Katsukawa}, {Lites}, {Nagata}, {Shibata}, {Shimizu}, {Shine}, {Suematsu},
  {Tarbell}, \& {Title}}]{Oka07}
{Okamoto}, T.~J., {et~al.} 2007, Science, 318, 1577

\bibitem[{{Sagdeev} \& {Galeev}(1969)}]{Sag69}
{Sagdeev}, R.~Z., \& {Galeev}, A.~A. 1969, {Nonlinear Plasma Theory}

\bibitem[{{Stone} {et~al.}(2008){Stone}, {Gardiner}, {Teuben}, {Hawley}, \&
  {Simon}}]{Sto08}
{Stone}, J.~M., {Gardiner}, T.~A., {Teuben}, P., {Hawley}, J.~F., \& {Simon},
  J.~B. 2008, \apjs, 178, 137

\bibitem[{{Suzuki} \& {Inutsuka}(2005)}]{Suz05}
{Suzuki}, T.~K., \& {Inutsuka}, S.-i. 2005, \apjl, 632, L49

\bibitem[{{Suzuki} \& {Inutsuka}(2006)}]{Suz06}
{Suzuki}, T.~K., \& {Inutsuka}, S.-I. 2006, Journal of Geophysical Research
  (Space Physics), 111, 6101

\bibitem[{{Tu} \& {Marsch}(1995)}]{Tu95}
{Tu}, C.-Y., \& {Marsch}, E. 1995, \ssr, 73, 1

\bibitem[{{van Ballegooijen} {et~al.}(2011){van Ballegooijen}, {Asgari-Targhi},
  {Cranmer}, \& {DeLuca}}]{Bal11}
{van Ballegooijen}, A.~A., {Asgari-Targhi}, M., {Cranmer}, S.~R., \& {DeLuca},
  E.~E. 2011, \apj, 736, 3

\bibitem[{{Velli}(1993)}]{Vel93}
{Velli}, M. 1993, \aap, 270, 304

\bibitem[{{Verdini} \& {Velli}(2007)}]{Ver07}
{Verdini}, A., \& {Velli}, M. 2007, \apj, 662, 669

\bibitem[{{Verdini} {et~al.}(2010){Verdini}, {Velli}, {Matthaeus}, {Oughton},
  \& {Dmitruk}}]{Ver10}
{Verdini}, A., {Velli}, M., {Matthaeus}, W.~H., {Oughton}, S., \& {Dmitruk}, P.
  2010, \apjl, 708, L116

\bibitem[{{Woolsey} \& {Cranmer}(2015)}]{Woo15}
{Woolsey}, L.~N., \& {Cranmer}, S.~R. 2015, \apj, 811, 136

\bibitem[{{Zank} \& {Matthaeus}(1992)}]{Zan92}
{Zank}, G.~P., \& {Matthaeus}, W.~H. 1992, Journal of Plasma Physics, 48, 85

\bibitem[{{Zhou} \& {Matthaeus}(1990)}]{Zho90}
{Zhou}, Y., \& {Matthaeus}, W.~H. 1990, \jgr, 95, 10291

\end{thebibliography}
\end{document}